\documentclass[%
preprint,
 nofootinbib,
 amsmath,amssymb,
 aps,
 a4paper
]{revtex4-1}
\pdfoutput=1

\usepackage{graphicx}% Include figure files
\usepackage{dcolumn}% Align table columns on decimal point
\usepackage{bm}% bold math
\usepackage{slashed}
\usepackage{color}
\usepackage{physics}

\usepackage{subfigure}
\begin{document}

%\preprint{XXXX}

\title{Non-thermal production of Higgsino dark matter by late-decaying scalar fields}

\author{Hajime Fukuda}
\email{hfukuda@hep-th.phys.s.u-tokyo.ac.jp}
\affiliation{Department of Physics, The University of Tokyo, Tokyo 113-0033, Japan}

\author{Qiang Li}
\email{qiangli@hep-th.phys.s.u-tokyo.ac.jp}
\affiliation{Department of Physics, The University of Tokyo, Tokyo 113-0033, Japan}

\author{Takeo Moroi}
\email{moroi@hep-th.phys.s.u-tokyo.ac.jp}
\affiliation{Department of Physics, The University of Tokyo, Tokyo 113-0033, Japan}

\author{Atsuya Niki}
\email{niki@hep-th.phys.s.u-tokyo.ac.jp}
\affiliation{Department of Physics, The University of Tokyo, Tokyo 113-0033, Japan}

\date{\today}% It is always \today, today,

\begin{abstract}

We study the non-thermal production of the Higgsino dark matter (DM). Assuming that the lightest neutral Higgsino is the lightest supersymmetric particle (LSP) in the minimal supersymmetric standard model, we calculate the relic abundance of the Higgsino LSP produced by the decay of late-decaying scalar field. In the calculation of the relic abundance, we have properly included the effects of coannihilation as well as the non-perturbative effect (known as the Sommerfeld effect). Contrary to the case of the thermal-relic scenario, in which the observed DM abundance is realized with the Higgsino mass of $\sim 1.2\ {\rm TeV}$, Higgsino DM is possible with a lighter Higgsino mass as the reheating temperature becomes lower than the Higgsino mass. The reheating temperature relevant for realizing the correct DM density is presented as a function of the Higgsino mass. 

\end{abstract}

\maketitle

%\tableofcontents

\section{Introduction}

Supersymmetry is one of attractive concepts in particle physics. The minimal supersymmetric standard model (MSSM) ameliorates the hierarchy problem in the standard model\,\cite{Feng:2013pwa} and makes it possible to unify the standard model gauge couplings at high energy\,\cite{Dimopoulos:1981zb}. It also provides a candidate for the dark matter (DM) in the universe assuming $R$-parity conservation. In the MSSM, supersymmetric particles are odd under $R$-parity whereas the standard model particles are even. Thus, the lightest supersymmetric particle (LSP) is stable, and the DM candidate\,\cite{Farrar:1978xj,Jungman:1995df}. 

Among supersymmetric particles, the lightest Higgsino is a good candidate for the DM.  Higgsinos are the supersymmetric partner of the Higgs bosons in the MSSM and are $\text{SU}(2)_L$ doublets with the hypercharge $Y=\pm\frac{1}{2}$. Assuming that the supersymmetric Higgs mass parameter, so-called the $\mu$-parameter, is smaller than the masses of other superparticles, the lightest Higgsino component becomes the LSP; in this paper, we concentrate on such a case.  After the electroweak symmetry breaking, the charged component becomes heavier than neutral ones mainly due to the radiative correction. Also, the Higgsino mixes with the bino and wino, the supersymmetric partners of $\text{U}(1)_Y$ and $\text{SU}(2)_L$ gauge bosons, respectively. As a result, two Majorana neutral fermions show up as mass eigenstates; the lightest one can be the LSP and plays the role of DM. 
From a bottom-up approach, Higgsino-like DM is one of the simplest DM models; it mainly interacts with the standard model gauge interaction. Its primary characteristic is its mass, making it an embodiment of the minimal DM scenario\,\cite{Cirelli:2005uq,Cirelli:2007xd}.

The most popular scenario realizing the Higgsino DM is the thermal relic scenario, in which thermally produced Higgsinos in the early universe are the origin of DM. The thermal relic scenario is simple and predictive. In particular, in order to realize the observed DM density ($\Omega_{\rm{DM}} h^2=0.11907$ \cite{Planck:2018vyg}), where $\Omega_{\rm DM}$ is the density parameter of DM and $h$ is the Hubble constant in units of $100\ {\rm km/sec/Mpc}$, the Higgsino mass is required to be $\sim 1.2\ {\rm TeV}$\,\cite{Hisano:2006nn,Cirelli:2007xd}.  
However, the thermal relic scenario assumes the standard evolution of the universe at the cosmic time around the freeze-out of the DM particle, which may not be the case in various models. In particular, if there exists a long-lived scalar field which dominates the early universe, its decay in the early universe may significantly affect the freeze-out process of the DM particle. Such a scenario, called non-thermal DM production scenario \cite{Moroi:1999zb}, opens a possibility to realize the Higgsino DM with Higgsino mass smaller than $1.2\ {\rm TeV}$.
A scenario with lighter Higgsino DM may ameliorate the naturalness of the electroweak symmetry breaking within the MSSM\,\cite{Feng:2013pwa}.  In addition, because the Higgsino LSP with its mass of $\mathcal{O}(100)\ {\rm GeV}$ is being probed by the on-going ATLAS\,\cite{ATLAS:2022rme,ATLAS:2024umc} and CMS experiments\,\cite{CMS:2023mny}, the discovery of the Higgsino LSP at those experiments should shed light on the non-thermal Higgsino DM scenario. 

In this paper, we study the non-thermal DM production scenario. In order to precisely calculate the relic abundance of the Higgsino DM in the non-thermal production scenario, we should take into account several effects which may complicate the calculation of the relic abundance. First, there exist three mass eigenstates, i.e., the lightest neutral Higgsino (i.e., the DM candidate) $\chi^0_1$, second-lightest neutral Higgsino $\chi^0_2$, and the charged Higgsino $\chi^\pm_1$, thus the coannihilations among these significantly affect the relic abundance. In addition, because Higgsinos have electroweak quantum numbers, the Sommerfeld effect may enhance the annihilation cross sections of Higgsinos \cite{Hisano:2003ec, Hisano:2004ds, Hisano:2006nn}. The mass splittings among the mass eigenstates, which are model dependent, may also affect the relic abundance.  In this paper, we carefully take into account these effects and calculate the relic density of the lightest neutral Higgsino.  Although the non-thermal production of the Higgsino DM was considered before \cite{Han:2019vxi}, these effects (particularly, the Sommerfeld effect) have not been fully considered. Thus, in this paper, we carefully taken into account these issues and examine the production process of the Higgsino DM from the decay of long-lived scalar field.

The organization of this paper is as follows. 
In Sec.\ \ref{sec:higgsino}, we summarize the properties of the Higgsino DM.
In Sec.\ \ref{sec:nonthermal}, a brief overview of the scenario of non-thermal DM production is given.
In Sec.\ \ref{sec:abundance}, we discuss the calculation of the relic abundance of the Higgsino DM in the non-thermal DM production scenario.
Results of our numerical calculation are given in Sec.\ \ref{sec:result}.
Sec.\ \ref{sec:conclusion} is devoted for conclusions and discussion.

\section{Higgsino DM}
\label{sec:higgsino}

In this section, we review the properties of the Higgsino DM and summarize the current constraints, assuming that the lightest neutral Higgsino is the LSP. To make our argument simple, we assume that supersymmetric particles other than Higgsinos are so heavy that they are absent at the time of the freeze-out of Higgsinos from the thermal bath.  

The Higgsino is the supersymmetric partner of the Higgs boson. As there are two Higgs bosons in the MSSM, $H_u$ and $H_d$,
there are two Higgsinos, $\tilde{h}_u = (\tilde{h}^+_u, \tilde{h}^0_u)$
and $\tilde{h}_d = (\tilde{h}^0_d, \tilde{h}^-_d)$, respectively.
Before the electroweak symmetry breaking, these two Weyl fermions form an $\text{SU}(2)_L$ doublet Dirac fermion with $Y = 1 / 2$.
After the electroweak symmetry breaking,  the Higgsino mixes with the bino and wino.  The mass eigenstates of neutralinos and charginos are denoted as $\chi^0_i$ (with $i=1$, $2$, $3$, $4$) and $\chi^\pm_i$ (with $i=1$, $2$), respectively.  Because we are interested in the Higgsino-like mass eigenstates, $\chi^0_1$, $\chi^0_2$, and $\chi^\pm_1$ are of our main concern, where $\chi^0_1$, $\chi^0_2$ are the lighter two neutralino mass eigenstates and $\chi^\pm_1$ is the lightest chargino mass eigenstate.  Notice that $\chi^0_1$ is assumed to be the LSP. Mass eigenvalues of these particles are denoted as $m_{\chi^0_1}$, $m_{\chi^0_2}$, and $m_{\chi^\pm_1}$, respectively.  We also introduce the following mass-difference parameters:
\begin{align}
    \Delta m_{\chi^0} \equiv &\,  m_{\chi^0_2} - m_{\chi^0_1},
    \\
    \Delta m_{\chi^\pm} \equiv &\,  m_{\chi^\pm_1} - m_{\chi^0_1}.
\end{align}

At the tree level, the Lagrangian for the chargino and neutralino mass terms is given by\,\cite{Martin:1997ns}
\begin{align}
  \mathcal{L} \supset - \frac{1}{2} \left( \tilde{w}^0, \tilde{b}, \tilde{h}^0_d, \tilde{h}^0_u \right)^T \mathcal{M}_0 \left( \tilde{w}^0, \tilde{b}, \tilde{h}^0_d, \tilde{h}^0_u \right)
  - \left( \tilde{w}^-, \tilde{h}^-_d \right)^T \mathcal{M}_c \left( \tilde{w}^+, \tilde{h}^+_u \right) + \text{H.c.},
\end{align}
where $\tilde{w}^0$ and $\tilde{w}^\pm$ are the neutral and charged wino, respectively, $\tilde{b}$ is the bino, 
and $\mathcal{M}_0$ and $\mathcal{M}_c$ are the mass matrices for the neutralino and chargino, respectively.  The mass matrices are given by
\begin{align}
  \mathcal{M}_0 = \begin{pmatrix}
    M_1 & 0 & - m_Z s_W c_\beta & m_Z s_W s_\beta \\
    0 & M_2 & m_Z c_W c_\beta & - m_Z c_W s_\beta \\
    - m_Z s_W c_\beta & m_Z c_W c_\beta & 0 & - \mu \\
    m_Z s_W s_\beta & - m_Z c_W s_\beta & - \mu & 0
  \end{pmatrix},
\end{align}
and 
\begin{align}
  \mathcal{M}_c = \begin{pmatrix}
    M_2 & \sqrt{2} m_W s_\beta \\
    \sqrt{2} m_W c_\beta & \mu
  \end{pmatrix},
\end{align}
where $M_1$ and $M_2$ are the bino and wino masses, respectively, $m_W$ and $m_Z$ are the masses of the $W$ and $Z$ bosons, respectively,
$s_W$ and $c_W$ are the sine and cosine of the weak mixing angle, respectively, $s_\beta$ and $c_\beta$ are $\sin\beta$ and $\cos\beta$, respectively, and the angle $\beta$ is defined such that $\tan\beta$ is the ratio of the vacuum expectation values of the two Higgs bosons.
For the pure Higgsino-like DM, we assume $|M_1|, |M_2| \gg |\mu| \gg m_Z$. The mass of the neutral and charged Higgsinos are approximately given by
\begin{align}
  % TODO use correct value
  m_{\chi^0_1} &\simeq \mu - \frac{1}{2}m_Z^2 (1 + \sin2\beta) \left(\frac{s_W^2}{M_1} + \frac{c_W^2}{M_2}\right), \\
  m_{\chi^0_2} &\simeq \mu + \frac{1}{2}m_Z^2 (1 - \sin2\beta) \left(\frac{s_W^2}{M_1} + \frac{c_W^2}{M_2}\right), \\
  m_{\chi^\pm_1} &\simeq \mu -  m_Z^2 \sin 2\beta \frac{c_W^2}{M_2}.
\end{align}
Here, we assume $\mu$ to be real and positive, whereas $M_1$ and $M_2$ are real, to avoid the CP violation. 

In addition to the tree-level mass splitting, there are radiative corrections to the Higgsino masses. 
As a result, the charged component becomes heavier the neutral ones.
The mass difference between the charged Dirac component and the neutral one is given by\,\cite{Nagata:2014wma} 
\begin{align}
  \Delta m_{\chi^\pm, \text{rad}} &\simeq \frac{\alpha_2}{4\pi} m_{\chi_1^\pm} s_W^2 f\qty(\frac{m_Z}{m_{\chi_1^\pm}})\\
  &\simeq 356\,\text{MeV} \qty(1 - \frac{3 m_Z}{2\pi m_{\chi_1^\pm}}),
\end{align}
where
\begin{align}
    f(x) \equiv 2\int_0^1 dt (1 + t) \ln \qty(1 + \frac{x^2(1 - t)}{t^2}).
\end{align}
%where $m_{\tilde{h}^+}$ and $m_{\tilde{h}^0}$ are the masses of the charged and neutral components, respectively.
In total, the mass splittings between the lightest neutralino and the second lightest neutralino and the lightest chargino are well approximated as
\begin{align}
  \Delta m_{\chi^0} &\simeq m_Z^2 \left|\frac{s_W^2}{M_1} + \frac{c_W^2}{M_2}\right|, \\
  \Delta m_{\chi^\pm} &\simeq \Delta m_{\chi^\pm, \text{rad}} + \Delta m_{\chi^\pm, \text{tree}},
\end{align}
respectively, where
\begin{align}
  \Delta m_{\chi^\pm, \text{tree}} \equiv \frac{1}{2}\Delta m_{\chi^0} + \frac{1}{2}m_Z^2 \sin2\beta \left(\frac{s_W^2}{M_1} - \frac{c_W^2}{M_2}\right).
\end{align}
The Higgsino mass differences are thus determined by the MSSM parameters, $M_1$, $M_2$, $\mu$ and $\tan\beta$ and are highly model dependent. 

Experimental constraints on the Higgsino DM from the direct detection experiments and those from the collider experiments can be complementary to each other\,\cite{Fukuda:2017jmk}. The major contribution to the spin-independent (SI) direct detection cross section between the Higgsino DM and nucleons is given by the inelastic $Z$-boson exchange and the elastic Higgs boson exchange. The former, if exists, gives 
a too large cross section. To suppress the inelastic $Z$-boson exchange, the mass difference between the lightest neutralino and the second lightest neutralino should be larger than $\mathcal{O}(100)\,\text{keV}$\,\cite{Nagata:2014wma}. For the latter, the scattering amplitude is proportional to the coupling between the LSP and the Higgs boson, which is given by
\begin{align}
  \frac{\partial m_\text{LSP}}{\partial v} =  \frac{\Delta m_{\chi^0} }{v}\qty(1 + \sigma \sin 2\beta),
\end{align}
where $v \simeq 246\,\text{GeV}$ is the vacuum expectation value of the Higgs boson and $\sigma \equiv \text{sgn}(\frac{s_W^2}{M_1} + \frac{c_W^2}{M_2})$. Assuming the Higgsino is much heavier than nucleons, the cross section between the lightest Higgsino and, for example, a proton is given by
\begin{align}
  \sigma_{\text{SI}} \simeq \frac{m_p^2}{\pi} f_p^2,
\end{align}
where $m_p$ is the proton mass and the lightest Higgsino-proton coupling $f_p$ is given by
\begin{align}
  f_p = \frac{\Delta m_{\chi^0} m_p}{v^2 m_h^2} 
  \qty( 1 + \sigma\sin 2\beta ) \qty(\frac{2}{9} + \frac{7}{9}\sum_{q = u,d,s} f_{q}^p).
\end{align}
Here, the form factor $f_{q}^p$ is the matrix element of the scalar operator between the proton states, defined as
\begin{align}
  f_{q}^p = \frac{m_q}{m_p}\left\langle p | \bar{q}q | p \right\rangle.
\end{align}
For the heavier quarks, we have used the QCD trace anomaly to relate the their operators to the light quark operators\,\cite{Shifman:1978zn}.
Using the values of the form factors from the QCD lattice calculation\,\cite{Abdel-Rehim:2016won}, we can calculate the SI cross section between the Higgsino DM and the protons as
\begin{align}
\sigma_{\text{SI}} \simeq 8.8 \times 10^{-48}\,\text{cm}^2 
\times 
\qty(\frac{\Delta m_{\chi^0}}{\text{GeV}})^2 
\qty( 1 + \sigma\sin 2\beta )^2.
\end{align}
Compared to the current experimental bound\,\cite{LZ_TeVPa}, Higgsino DM with its mass of $\mathcal{O}(100)\,\text{GeV}$ and
mass differences of $\mathcal{O}(\text{GeV})$ is being tested.

On the other hand, for smaller mass differences, collider experiments can be effective to constrain the Higgsino DM. If $\mathcal{O}(100)\,\text{MeV} \gg \Delta m_{\chi^\pm, \text{tree}}$, $\Delta m_{\chi^\pm} \sim \Delta m_{\chi^\pm, \text{rad}}$ and the chargino mainly decay into the lightest neutralino and a pion. The chargino lifetime is relatively long\,\cite{Chen:1996ap,Thomas:1998wy,Ibe:2023dcu}, 
\begin{align}
  c\tau \sim 0.7\,\text{cm} \times \qty(\frac{350\,\text{MeV}}{\Delta m_{\chi^\pm}})^3 \qty(1 - \frac{m_{\pi^\pm}^2}{\Delta m_{\chi^\pm}^2})^{-1/2},
\end{align}
where $m_{\pi^\pm}$ is the charged pion mass. Therefore, the chargino can travel a macroscopic distance before decaying and it can be detected as a disappering track at the collider experiments\,\cite{Fukuda:2017jmk}. The current bound on the chargino mass is $\sim 200\,\text{GeV}$\,\cite{ATLAS:2022rme,CMS:2023mny}. If the mass difference is larger, $\Delta m_{\chi^\pm} \sim \Delta m_{\chi^0} \lesssim 1\,\text{GeV}$, the chargino lifetime is not long enough to be detected as a disappearing track at tracking detectors anymore, but the daughter pion can be detected as a soft displaced track\,\cite{Fukuda:2019kbp}. The current bound on the chargino mass in this case is also $\sim 170\,\text{GeV}$\,\cite{ATLAS:2024umc} for $\Delta m_{\chi^\pm} \sim \Delta m_{\chi^0} \sim 0.5\,\text{GeV}$.

Because the mass splittings among Higgsinos are model dependent, as explained so far, we consider several cases in our numerical calculation. Given the experimental constraints discussed above, the mass splittings we adopt are summarized in Table \ref{table:dm}.

\begin{table}[t]
    \begin{center}
    \begin{tabular}{r|rrr}
    \hline\hline
    & ~Model A & ~Model B & ~Model C
    \\
    \hline
    $\Delta m_{\chi^\pm}$ (GeV) & 0.35 & 0.5 & 1
    \\
    $\Delta m_{\chi^0}$ (GeV) & 0.1 & 0.5 & 1
    \\
    \hline\hline
    \end{tabular}
    \caption{The Higgsino mass splittings adopted in our numerical study. }
    \label{table:dm}
    \end{center}
\end{table}

\section{Non-thermal production of DM}
\label{sec:nonthermal}

In this section, we introduce the scenario of the non-thermal DM production. (For earlier works, see Refs.\ \cite{Moroi:1999zb, Jeannerot:1999yn, Giudice:2000ex, Fujii:2001xp, Fornengo:2002db, Gelmini:2006pw, Gelmini:2006pq, Acharya:2009zt, Arcadi:2011ev, Moroi:2013sla, Kane:2015qea, Drees:2018dsj, Han:2019vxi}.) Particularly, we consider the case that the universe was once dominated by a late-decaying scalar field, which we call $\phi$, and that its decay contributes to the DM production. The candidates of the late-decaying scalar field include the moduli fields which show up in the string-inspired models or saxion field which is a real and massive scalar field in association with the axion. Although we do not specify the model behind $\phi$ and our following discussion holds irrespective of its detailed origin, we call $\phi$ as a modulus field. In the following, we adopt the following assumptions:
\begin{itemize}
\item $\phi$ is very weakly interacting and that the interaction of $\phi$ with standard-model fields are suppressed by the reduced Planck scale $M_{\rm Pl}$.
\item The mass of $\phi$, which is expected to be of the order of the SUSY breaking mass scale, is much smaller than the Planck scale. Then, because the interaction of the modulus field is expected to be very weak, the potential of $\phi$ is well approximated by a parabolic one:
\begin{align}
  V = \frac{1}{2} m_\phi^2 \phi^2,
\end{align}
choosing $\phi=0$ as the minimum of the potential.
\item The modulus $\phi$ decays only into visible sector fields (including the DM particle).
\end{itemize}
We parameterize the decay rate of $\phi$ as 
\begin{align}
  \Gamma_\phi = \frac{\kappa}{4\pi} \frac{m_\phi^3}{M_{\rm Pl}^2},
  \label{Gammaphi}
\end{align}
where $\kappa$ is a free parameter and $M_{\rm Pl}\simeq 2.4\times 10^{18}\ {\rm GeV}$ is the reduced Planck scale. For the moduli field in string theory, $\kappa\sim O(1)$ is expected. For other long-lived fields, such as the saxion, the cut-off scale giving rise to the dimension-5 operator given in Eq.\ \eqref{Gammaphi} may be lower than the reduced Planck scale, which corresponds to $\kappa\gg O(1)$.
Our following analysis is applicable to both cases with long-lived scalar fields which once dominate the universe and we do not exclude the latter possibility.

The amplitude of the modulus field may be displaced from the minimum of the potential in the very early universe so that it may significantly affect the evolution of the universe. The equation of motion of $\phi$ is given as
\begin{align}
  \ddot{\phi} + 3 H \dot{\phi} + \frac{\partial V}{\partial \phi} \simeq 0,
  \label{EoM}
\end{align}
where the ``dot'' denotes the derivative with respect to time, $H\equiv \dot{a}/a$ (with $a$ being the scale factor) is the expansion rate of the universe, and $V$ is the scalar potential. Here, the effect of the decay of $\phi$ is neglected; Eq.\ \eqref{EoM} is applicable when $H\gg\Gamma_\phi$. 

Now, we consider the evolution of the modulus field $\phi$, assuming that its initial amplitude is non-vanishing.  In the early universe when $H\gg m_\phi$, the Hubble-friction term (i.e., the second term in the left-hand side of Eq.\ \eqref{EoM}) is so effective that $\phi$ slowly rolls toward the minimum of the potential. Then, with the expansion of the universe, the expansion rate becomes decreased. Once $H$ becomes comparable to $m_\phi$, the modulus field starts to oscillate around the minimum of the potential. After such an epoch, it is convenient to define the energy density of the modulus as
\begin{align}
  \rho_\phi \equiv \left\langle \frac{1}{2} \dot{\phi}^2 + \frac{1}{2} m_\phi^2 \phi^2 \right\rangle_{\rm osc},
\end{align}
with $\langle\cdots\rangle_{\rm osc}$ being the oscillation average, to study the evolution of the modulus field.  The evolution of $\rho_\phi$ is governed by the following equation:
\begin{align}
  \dot{\rho}_\phi + 3 H \rho_\phi = -\Gamma_\phi \rho_\phi,
  \label{dot(rhophi)}
\end{align}
thus $\rho_\phi$ behaves as
\begin{align}
  \rho_\phi (t) \simeq \rho_\phi (t_0) \left[ \frac{a(t)}{a(t_0)} \right]^{-3} e^{-\Gamma_\phi t}.
\end{align}
Using the fact that $H\sim O(t^{-1})$, we can see that the energy density of the modulus behaves as that of non-relativistic matter when $\Gamma_\phi\lesssim H\lesssim m_\phi$, which implies that the energy density of $\phi$ decreases more slowly than that of radiation.
Accordingly, the modulus dominates the universe if its initial amplitude is large enough. Hereafter, we consider such a case; then, the following argument is insensitive to the initial amplitude of the modulus field as far as the modulus once dominates the universe. Then, the modulus decays when $H$ becomes comparable to $\Gamma_\phi$ and the energy density of the modulus is converted to that of radiation. We call such an epoch as the reheating epoch. The cosmic temperature at the time of the reheating can be estimated by using the relation $H\sim \Gamma_\phi$; in our convention, the reheating temperature $T_{\rm R}$ satisfies
\begin{align}
    \frac{\pi^2}{30} g_* (T_{\rm R}) T_{\rm R}^4 
    = 3 M_{\rm Pl}^2 \Gamma_\phi^2,
  \label{Tr}
\end{align}
where $g_*$ is the effective number of the massless degrees of freedom for the calculation of the energy density. In our following discussion, $T_{\rm R}$ (instead of $\Gamma_\phi$ or $\kappa$) is treated as a free parameter.

We are interested in the case that the modulus can decay into a pair of DM (as well as into standard-model particles). The evolution of the number density of DM is governed by 
\begin{align}
  \dot{n}_{\rm DM} + 3 H n_{\rm DM} = - \langle \sigma v_{\rm rel} \rangle 
  \left( n_{\rm DM}^2 - \bar{n}_{\rm DM}^2 \right) + B_{\rm DM} \Gamma_\phi n_\phi,
  \label{dot(ndm)}
\end{align}
where $\langle\sigma v_{\rm rel} \rangle$ is the thermally averaged pair-annihilation cross section, $B_{\rm DM}$ is the averaged number of the DM particles produced by the modulus decay, 
\begin{align}
  n_\phi \equiv \frac{\rho_\phi}{m_\phi},
\end{align}
and $\bar{n}_{\rm DM}$ is the thermal-equilibrium value of the DM number density. Furthermore, the evolution of the radiation sector is governed by the following equation:
\begin{align}
    \dot{s}_{\rm rad} + 3 H s_{\rm rad} = 
    \frac{1}{T} \left[
    (1-\bar{B}) \Gamma_\phi \rho_\phi 
    + 2 \langle E_{\rm DM} \rangle
    \langle \sigma v_{\rm rel} \rangle 
    \left( n_{\rm DM}^2 - \bar{n}_{\rm DM}^2 \right)
    \right],
    \label{dot(srad)}
\end{align}
where $s_{\rm rad}$ is the entropy density of radiation,  $\langle E_{\rm DM}\rangle$ (which is taken to be $m_{\rm DM}+\frac{3}{2} T$ in our numerical calculation) is the thermally averaged energy of DM particle, and 
\begin{align}
\bar{B} \equiv \frac{\langle E_{\rm DM}\rangle B_{\rm DM}}{m_\phi}.
\end{align}
The cosmic temperature $T$ is related to $s_{\rm rad}$ as
\begin{align}
    s_{\rm rad} (T) = \frac{2\pi^2}{45} h_* (T) T^3,
\end{align}
where $h_*$ is the effective number of the massless degrees of freedom for the calculation of the entropy density. With the cosmic temperature being given, we can calculate the energy density of radiation as
\begin{align}
    \rho_{\rm rad} (T) = \frac{\pi^2}{30} g_* (T) T^4.
\end{align}
For the case of our interest, the evolution of the DM density can be obtained by simultaneously solving Eqs.\ \eqref{dot(rhophi)}, \eqref{dot(ndm)} and \eqref{dot(srad)} with using
\begin{align}
  H = \sqrt{\frac{\rho_\phi + \rho_{\rm rad}+\rho_{\rm DM}}{3 M_{\rm Pl}^2}},
\end{align}
where $\rho_{\rm DM}$ is the energy density of the DM.

If annihilation processes of the DM with other particles (i.e., so-called coannihilation processes) are relevant, they can be embedded into the effective annihilation cross section $\langle \sigma_{\rm{eff}} v_{\rm rel}\rangle$. Assuming the chemical equilibrium among the particles participating to the coannihilation processes, we obtain
\begin{align}
	\langle \sigma_{\rm{eff}} v_{\rm rel}\rangle&=\sum_{ij}\langle \sigma_{ij} v\rangle 
    \frac{g^2_{\rm{eff}}(x=\infty)}{g^2_{\rm{eff}}(x)}
    (1+\triangle_i)^{3/2}(1+\triangle_j)^{3/2}\exp[-x(\triangle_i+\triangle_j)],
    \label{<sigmav>}
\end{align}
where $\langle \sigma_{ij}v\rangle$ denotes the thermally averaged cross of $i$- and $j$-th particles:
\begin{align}
	\langle \sigma_{ij}v\rangle&=\left(\frac{m_{\rm DM}}{4\pi T}\right)^{3/2}\int 4\pi v^2dv (\sigma_{ij} v) \exp\left(-\frac{m_{\rm DM} v^2}{4T}\right),
\end{align}
with $\sigma_{ij}$ being the cross section of the annihilation process. Here $\triangle_i\equiv (m_i-m_{\rm DM})/m_{\rm DM}$ (with $m_i$ being the mass of $i$-th particle),
$g_{\rm eff}$ is the effective number of the spin degrees of freedom, and
\begin{align}
    x\equiv \frac{m_{\rm DM}}{T}.
\end{align} 

In the scenario of the non-thermal production of the DM, there are two production processes of DM in the early universe; one is the scattering processes in the thermal bath and the other is the decay of the modulus field. In particular, if the reheating temperature is of the order of the freeze-out temperature of the DM particle or lower, the contribution from the modulus decay may become sizable; in such a case, the relic abundance of the DM can be altered compared to the ordinary thermal relic scenario, resulting in a significant change of the DM parameters explaining the present mass density of the DM. Notice that the DM particles produced before the reheating epoch are diluted by the entropy production due to the modulus decay, and hence the DM density in the present universe is primarily determined by the total amount of the DM produced at the time of the reheating (or later).  The dependence of DM abundance on $B_{\rm DM}$ is understood as follows:
\begin{itemize}
\item When $B_{\rm DM}$ is small enough, (almost) all the DM particles from the modulus decay survive. In such a case, the DM density in the present universe is approximately proportional to $B_{\rm DM}$.
\item With sizable $B_{\rm DM}$, DM particles produced by the modulus decay is so large that a significant fraction of DM particles experience the pair annihilation process; the pair annihilation proceeds until the annihilation rate becomes comparable to the expansion rate of the universe. In this case, the present DM density becomes insensitive to $B_{\rm DM}$.
\end{itemize}
Based on the above arguments, the present number density of DM is approximately expressed as \cite{Moroi:1999zb}
\begin{align}
  n_{\rm DM} (T_{\rm now}) \sim
  \frac{s (T_{\rm now})}{s (T_{\rm R})} \times \mbox{min}
  \left( \frac{\Gamma_\phi}{\langle \sigma v \rangle}, \frac{B_{\rm DM}\Gamma_\phi^2 M_{\rm Pl}^2}{m_\phi} \right),
  \label{relic estimate}
\end{align}
where $s(T)$ is the entropy density of the universe as a function of the cosmic temperature $T$, and $T_{\rm now}\simeq 2.7\, {\rm K}$ is the present cosmic temperature.  Using the relation between the reheating temperature $T_{\rm R}$ and the decay rate, we can see that, when $B_{\rm DM}\gtrsim B_*$, the annihilation of the DM becomes so efficient at the time of the reheating that the first term in the right-hand side of Eq.\ \eqref{relic estimate} dominates over the second term; using the relation between the reheating temperature $T_{\rm R}$ and the decay rate given in Eq.\ \eqref{Tr}, $B_*$ is estimated as
\begin{align}
    B_*\sim \frac{m_\phi}{M_{\rm Pl} T_{\rm R}^2\langle \sigma v \rangle}
    \sim
    10^{-5} \times
    \left( \frac{m_\phi}{100\ {\rm TeV}} \right)
    \left( \frac{m_{\rm DM}}{1\ {\rm TeV}} \right)^{2}
    \left( \frac{T_{\rm R}}{1\ {\rm GeV}} \right)^{-2},
\end{align}
where, in the second equality, we used the estimation of the Higgsino annihilation cross section.

Although the above argument provides an order-of-magnitude estimation of the DM relic density, more careful analysis is needed for a precise prediction of the DM abundance. For this purpose, in the following, we solve the Boltzmann equations numerically and calculate the relic abundance of the Higgsino DM for various choices of model parameters.

\section{Calculation of the DM abundance}
\label{sec:abundance}

\subsection{Sommerfeld effects}

Sommerfeld effect is a non-perturbative quantum effect caused by exchange of light mediators. Particularly, for the study of the DM relic abundance, the Sommerfeld effect may significantly enhance or suppress the annihilation cross section of DM pairs so it may play an important role \cite{Hisano:2003ec, Hisano:2004ds, Hisano:2006nn}.  For the case of DM candidates having electroweak quantum numbers, like the wino or Higgsino in the supersymmetric model or $SU(2)_L$ quintuplet, the Sommerfeld effect is known to significantly alter the thermal relic abundance. Because the Higgsino multiplet contains several mass eigenstates which have relatively small mass splittings, as explained in section \ref{sec:higgsino}, we should take into account the coannihilations of neutralinos $\chi_1^0$, $\chi_2^0$ and charginos $\chi_1^{\pm}$. 

In order to study the Sommerfeld effect on the annihilation rates, we derive the effective actions for the non-relativistic two-body states built from pairs of $\chi_1^0$, $\chi_2^0$ and  $\chi_1^{\pm}$. Following the procedure in Ref.\ \cite{Hisano:2004ds}, we firstly integrate out the gauge bosons in the MSSM which give the interaction action between currents of Higgsinos. Then, the high momentum modes are integrated out, inducing the absorptive parts in  the non-relativistic effective action. 
The final effective action has the form of 
\begin{equation}
    S^{(2)}=\int d^4x d^3r \sum_{S,Q}\Phi^\dagger_{S,Q}(x,\vec{r})\left[\left(i\partial_{x^0}+\frac{\nabla^2_x}{4m_{\chi_1^0}}+\frac{\nabla^2_r}{m_{\chi_1^0}} \right)-\mathcal{H}_Q^S(r)+2i\Gamma_Q^S\delta(r)
    \right]\Phi_{S,Q}(x,\vec{r}),
    \label{effaction}
\end{equation}
where $\Phi$ are vectors of composite field, the argument $x$ is the center-of-mass coordinate of the two-body states and $\vec{r}$ is the relative coordinate vector. The components of $\Phi$ are built from two-body states with the same spin $S$ and charge $Q$. The potential $\mathcal{H}(r)$  describes the Coulomb and Yukawa forces given by the exchange of gauge bosons.  The absorptive part $\Gamma$ corresponds to short-distance annihilation of two-body states, can be obtained by calculating diagrams of transitions between two-body states and using the optical theorem. We calculate $\Gamma$ with taking into account all the relevant box, triangle, and self-energy diagrams. The formulas for $\mathcal{H}(r)$ and $\Gamma$ are shown in the appendix.

From the effective action in Eq.\eqref{effaction}, one can derive the Schwinger-Dyson equation of the Green function $\langle0|T\Phi(x,\vec{r})\Phi^\dagger(y,\vec{r}')|0\rangle$ which can be decomposed into different partial wave mode. Throughout our analysis, we focus on the s-wave annihilation processes. Hence, at the leading order in $\Gamma$,  for the calculation of the Sommerfeld factor, we should solve the multi-channel Schrodinger equation of the following form:
\begin{equation}
	-\frac{1}{2 m_{\rm r}}\frac{d^2}{dr^2}\Psi(r)+\mathcal{H}(r)\Psi(r)=E\Psi(r),
 \label{scheq}
\end{equation}
where $\Psi$ is a matrix and consists of wave functions for transitions between two-body states with the same $Q$ and $S$, $m_{\rm r}$ is the reduced mass of the lightest two-particle state, and $E$ is the energy in the center-of-mass frame. We parameterize $E$ as
\begin{align}
    E = \frac{1}{2} m_{\rm r} v_*^2,
\end{align}
with $v_*$ being the relative velocity between particles in the lightest two-particle state. For heavier states (with the reduced mass $m_r^{(h)}$), the relative velocity $v_{\rm{rel}}$ is related to $v_*$ as
\begin{align}
    \frac{1}{2} m_{\rm r} v_*^2 = 
    \frac{1}{2} m_r^{(h)}v_{\rm{rel}}^2  + \delta M,
\end{align}
where $\delta M$ is the mass differences between heavier and lightest two-body states. For Higgsinos, we have four different channels with different values of $(Q,S)$, so we solve four multi-channel Schrodinger equations in total. 

In order for the calculation of the Sommerfeld factor, we need a solution of Eq.\ 
\eqref{scheq} such that it satisfies $\Psi(r=0)=\mathbf{1}$ and that, at $r\rightarrow \infty$, it has the form of an out-going wave \cite{Hisano:2004ds}. Because the Yukawa potentials in $\mathcal{H}(r)$ vanish at very large $r$, each two-body wave function $\Psi_{ij}$ (with indices $i, j$ labeling the two-body states) obeys the differential equation of the following general form at $r\rightarrow\infty$:
\begin{equation}
\frac{d^2}{dr^2} \Psi_{ij}(r)+\frac{\alpha'}{r}\Psi_{ij}(r)+\beta \Psi_{ij}(r) = 0,
\label{1dschro}
\end{equation}
where $\alpha'$ takes the values of either $2m_r\alpha $ or 0, and $\beta$ is either $2m_r(E-\delta M)$ or $2m_rE$, depending on the specific channels involved. The solution is given by
\begin{equation}
    \Psi_{ij}(r\rightarrow\infty) = 
	d_{ij} \left[re^{-r\sqrt{-\beta}}\, U(1-\frac{\alpha'}{2\sqrt{-\beta}}, 2, 2r\sqrt{-\beta}) \right]^*,
 \label{outgoing}
\end{equation}
where $d_{ij}$ is a constant, and $U$ is Tricomi's confluent hypergeometric function. In our numerical calculation, we first impose the boundary condition consistent with Eq.\ \eqref{outgoing} at $r\sim\infty$ (i.e., large enough value of $r$), and solve Eq.\ \eqref{scheq} from $r\sim\infty$ to $r=0$; let us call the solution obtained with this procedure as $\hat{\Psi}(r)$. In general, $\hat{\Psi}(r)$ does not satisfy the relevant boundary condition at $r=0$; the function $\Psi$ of our interest can be obtained as $\Psi(r)=\hat{\Psi}(r) \hat{\Psi}^{-1}(0)$. 

After obtaining numerical solution of $\Psi$'s, we can evaluate the  Sommerfeld enhanced annihilation cross sections. For any two-particle state $i$ with charge $Q$, the result is given by
\begin{align}
\sigma_{i\rightarrow \rm{light}}v_{\rm{rel}}&=\sum_{S}c_i\times s_i\times \Psi_{ij}^\infty(\Gamma_Q^S)_{jk}\Psi^{\infty *}_{ik}
\end{align}
where $\Psi^\infty$ is the asymptotic value of $\Psi$ at $r \rightarrow \infty$, $c_i=2$ ($c_i=1$) if $i$ is made of two identical (different) particles, and $s_i=1$ ($s_i=3$) for the annihilation of state $i$ with the spin configuration  $S=0$ ($S=1$).
%for Winos:
%\begin{align}
%	\sigma_{\chi_0 \chi_0 \rightarrow \rm{light}}&=2\times \Psi_{2i}(\Gamma_0^{S=0})_{ij}\Psi^*_{2j}\\
%	\sigma_{\chi^+ \chi^-\rightarrow \rm{light}}&=\Psi_{1i}(\Gamma_0^{S=0})_{ij}\Psi^*_{1j} + 3\cdot  \Psi_{11}\Gamma_0^{S=1}\Psi^*_{11}\\
%	\sigma_{\chi_0\chi^-\rightarrow \rm{light}}&= \Psi_{11}\Gamma_1^{S=0}\Psi^*_{11} + 3\cdot  \Psi_{11}\Gamma_1^{S=1}\Psi^*_{11}\\
%	\sigma_{\chi^-\chi^-\rightarrow \rm{light}}&= 2\times \Psi_{11}\Gamma_2^{S=0}\Psi^*_{11} 
%\end{align}

We take into account the running effect of gauge coupling constants. Because the Sommerfeld effect is due to long-distance effect, we use the gauge coupling constants at $Q=m_Z$ (with $Q$ being the renormalization group scale) in solving the Schrodinger equations. On the other hand, the tree-level annihilation cross sections, which are given by the absorptive parts, should be evaluated at Higgsino mass scale $Q=|\mu|$.  In our analysis, we use the one-loop renormalization group equation to take into account the scale dependence of gauge coupling constants:
\begin{equation}
	\frac{d}{d \ln Q}\alpha_a^{-1}=-\frac{b_a}{2\pi},
\end{equation}
where $\alpha_a=g_i^2/4\pi$ (with $a=1$ and $2$ for $U(1)_Y$ and $SU(2)_L$, respectively).  Because the renormalization group running below the mass scale of superparticles are important for our study, we use the standard-model renormalization group equations, i.e., $(b_1,b_2)=(41/6,-19/6)$.

\subsubsection{Sommerfeld enhancement }

The Sommerfeld enhancement factor for any channel $i$ is defined as the enhanced cross annihilation section normalized by the tree-level result:
\begin{equation}
	S_i=\frac{ \sigma_{i\rightarrow \rm{light}} v_{\rm{rel}}}{c_i \Gamma_{ii}},
\end{equation}
where the label $i$ refers to two-body states.

\begin{figure}[t]
	\centering
	\includegraphics[width=0.75\textwidth]{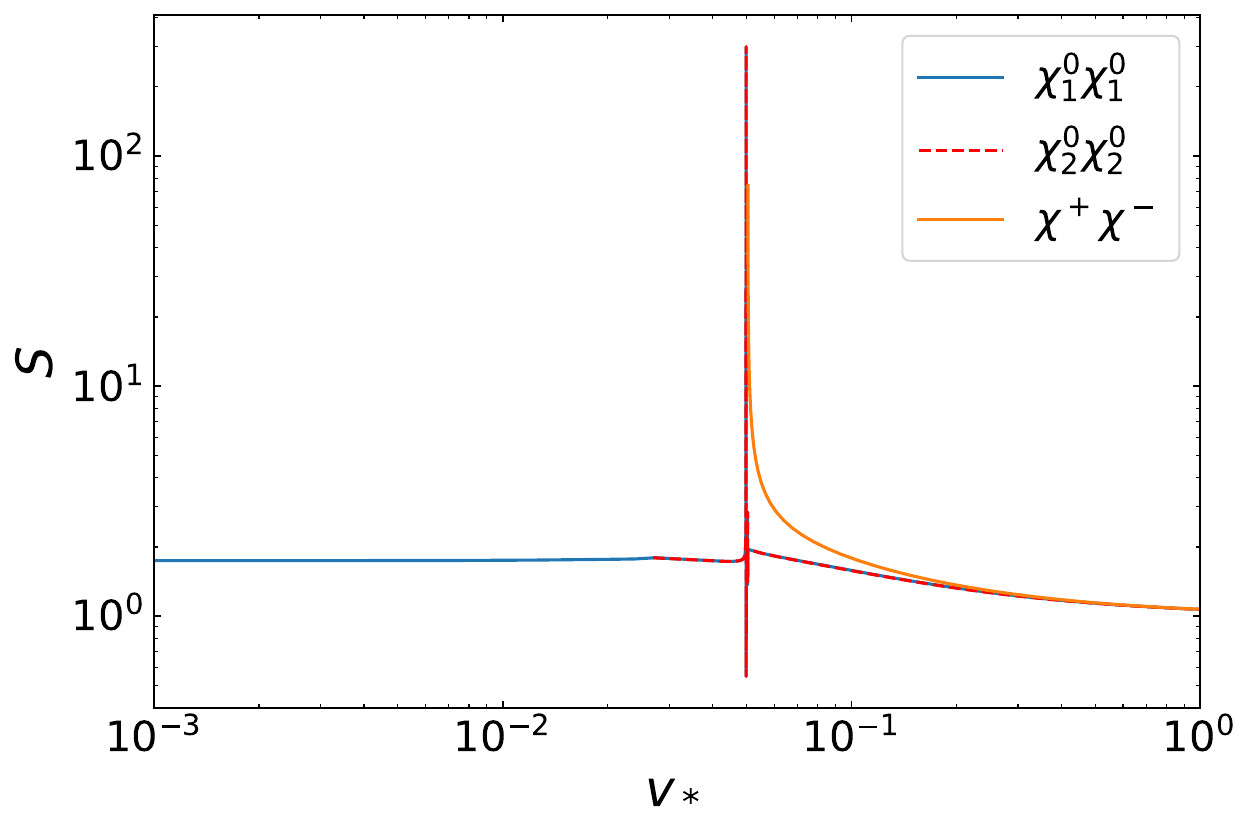}
	\caption{The Sommerfeld enhancement factors for the $^1S_0$ partial-wave annihilation of $\chi_1^0\chi_1^0$, $\chi_2^0\chi_2^0$ and $\chi^+\chi^-$ channels. $v_*$ is the relative velocity between $\chi_1^0$s in the lightest two-particle state $\chi_1^0\chi_1^0$. For other heavier channels, the relative velocity $v_{\rm{rel}}$ is related to $v_*$ by the mass splittings.}
	\label{fig:0v3}
\end{figure}

\begin{figure}[t]
	\centering
	\includegraphics[width=0.75\textwidth]{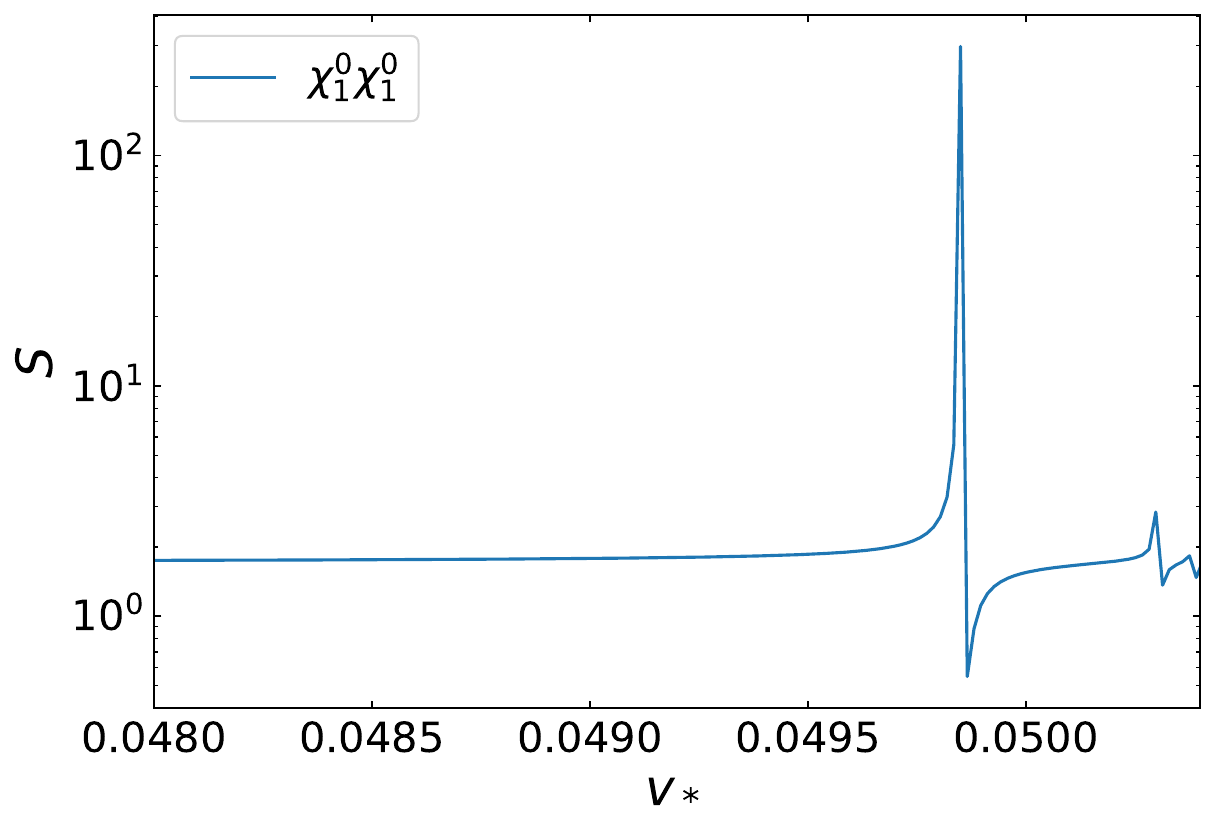}
	\caption{The first three peaks in Coulomb resonances of Sommerfeld factors of $\chi_1^0\chi_1^0$  through $^1S_0$ partial-wave annihilation. For the mass splitting $\Delta m_{\chi^\pm}=0.35\rm{GeV}$, the threshold for $\chi^+\chi^-$ is $v_*\simeq 0.0505$. }
	\label{fig:0v3-reso}
\end{figure}

In order to show the typical behavior of the Sommerfeld factors, in Fig.\ \ref{fig:0v3}, we plot the Sommerfeld factors for $Q=0$ states $\chi_1^0\chi_1^0$, $\chi_2^0\chi_2^0$ and $\chi^+\chi^-$ through the $^1S_0$ partial-wave annihilation as functions of $v_*$.  Here, we take $m_{\chi_0^1}=1.1\ {\rm TeV}$ and adopt the mass splittings of Model A given in Table \ref{table:dm} (i.e., $\Delta m_{\chi^\pm}=0.35\ {\rm GeV}$ and $\Delta m_{\chi^0}=0.1\ {\rm GeV}$), for which the channel $\chi_2^0\chi_2^0$ opens at $v_*\simeq 0.02697$, and the threshold for appearance of $\chi^+\chi^-$ in the asymptotic state is $v_*\simeq 0.0505$. At small velocities, the Sommerfeld factor approaches to $\sim 1.715$ and velocity-independent, which is a typical behavior given by Yukawa potential. On the other hand, at large velocity, the perturbative picture becomes relevant and the Sommerfeld factor approaches to $1$. The behavior of the Sommerfeld factor near the threshold is rather complicated.  Just above the threshold of $\chi^+\chi^-$, the on-shell charged Higgsinos state can be produced and the enhancement factor of $\chi^+\chi^-$ diverges as the inverse of relative velocity between them $1/v^{\chi^\pm}_{\rm{rel}}$. Such divergence is expected because at large distance, the Coulomb potential dominates the interactions. In this regime, an analytic estimate to the Sommerfeld factor of $\chi^+\chi^-$ is given by \cite{Landau:1991wop}
\begin{equation}
	S\simeq \frac{\pi \alpha/v^{\chi^\pm}_{\rm{rel}}}{1-e^{-\pi\alpha/v^{\chi^\pm}_{\rm{rel}}}}.
\end{equation}
Just below the $\chi^+\chi^-$ threshold, there exist peaks in the behavior of the Sommerfeld factor. The detailed behavior of the Sommerfeld factor just below the threshold is shown in Fig.\ \ref{fig:0v3-reso}. The pattern of peaks matches the binding energies of Coulomb potential 
\begin{equation}
	\frac{m_{\chi_1^0}v_*^2}{4}=2\Delta m_{\chi^\pm}-\frac{m_{\chi_1^0}}{4}\frac{\alpha^2}{n^2},
\end{equation} 
with $n=1,2,\cdots$. At these locations of $v_*$, the bound states of charged Higgsinos appears, leading to the resonance peaks in the Sommerfeld factors. 

\subsubsection{Thermally averaged cross sections}

With the Sommerfeld enhanced cross sections of all the channels, we calculate  the thermally averaged cross section given in Eq.\ \eqref{<sigmav>}. For the case of the Higgsino DM, $\chi^0_1$, $\chi^0_2$, and $\chi^\pm_1$ are the particles relevant for the study of the relic abundance.  Then, $g_{\rm eff}$ is given by
\begin{equation}
	g_{\rm{eff}}=2+2(1+\Delta m_{\chi^0}/m_{\chi_1^0})^{3/2}e^{-x\Delta m_{\chi^0}/m_{\chi_1^0}}+4(1+\Delta m_{\chi^\pm}/m_{\chi_1^0})^{3/2}e^{-x\Delta m_{\chi^\pm}/m_{\chi_1^0}},
\end{equation}
where $x=m_{\chi_1^0}/T$.

\begin{figure}[t]
	\centering
	\includegraphics[width=0.75\textwidth]{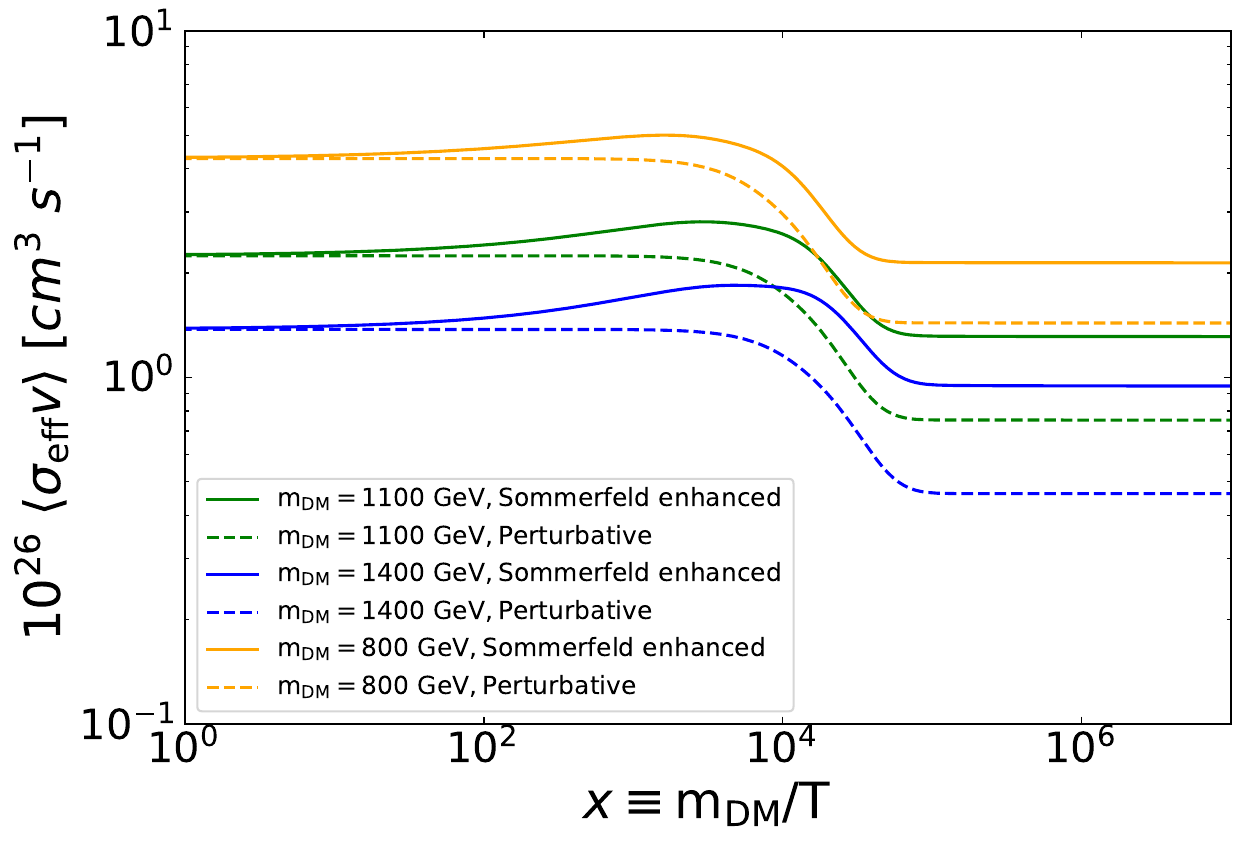}
	\caption{The perturbative and Sommerfeld enhanced thermally averaged cross sections $\langle \sigma v_{\rm{rel}}\rangle$ as functions of $x=m_{\chi_1^0}/T$, taking $m_{\chi^0_1}=0.8$, $1.1$, and $1.4\ {\rm TeV}$.   
    Here, mass-splitting parameters of model A are adopted (i.e., $\Delta m_{\chi^\pm}=0.35\ {\rm GeV}$ and $\Delta m_{\chi^0}=0.1\ {\rm GeV}$).}
	\label{fig:thermal}
\end{figure}

In Fig.\ \ref{fig:thermal}, we plot $\langle \sigma v_{\rm{rel}}\rangle$ as a function of $x$ with and without including the Sommerfeld effect, adopting the mass-splitting parameters of model A. For the chosen values of mass splittings, the decoupling of heavier neutralino and charginos occurs at around $x\sim 10^4$. After the decoupling, the enhancement of $\langle \sigma v_{\rm{rel}}\rangle$ becomes constant, due to the velocity-independent Sommerfeld factor of $\chi_1^0\chi_1^0$ at very low velocities as shown in Fig.\ \ref{fig:0v3}. 

\subsection{Relic abundances}
To obtain the relic abundance of Higgsino DM which has the mass $m_{\rm{DM}}=m_{\chi_1^0}$, we numerically solve the Boltzmann equations \eqref{dot(rhophi)}, \eqref{dot(ndm)} and \eqref{dot(srad)}. For this purpose, it is convenient to use the following  dimensionless quantities to rewrite the Boltzmann equations:
\begin{align}
	\Phi\equiv \frac{\rho_{\phi} a^3}{T_{\rm{R}}}, R\equiv \rho_{R}a^4, X\equiv n_{\rm{DM}} a^3, A=aT_{\rm{R}}.
\end{align}
Then the evolution equations become\,\cite{Drees:2017iod,Kane:2015qea}
\begin{align}
	\tilde{H}\frac{d\Phi}{dA}&=-c_\rho^{1/2}A^{1/2}\Phi, \\
	\tilde{H}\frac{dX}{dA}&=\frac{c_{\rho}^{1/2}T_RB_{\rm{DM}}}{m_\phi}+\sqrt{3}M_{\rm{Pl}}T_RA^{-5/2}\langle\sigma_{\rm{eff}}v_{\rm rel}\rangle (X^2_{\rm{EQ}}-X^2), \\
	\frac{dT}{dA}&=\left(1+\frac{T}{3h_*}\frac{dh_*}{dT}\right)^{-1}\biggl[
		-\frac{T}{A}+\frac{15T_R^6}{2\pi^2\sqrt{3}M_{\rm{Pl}}HT^3h_*A^{11/2}}\biggl(c_\rho^{1/2}A^{3/2}(1-\bar{B})\Phi \nonumber\\
		   &+\sqrt{3}M_{\rm{Pl}}\frac{2\langle E_{\rm{DM}}\rangle \langle\sigma_{\rm{eff}}v_{\rm rel}\rangle }{A^{3/2}} (X^2-X_{\rm{EQ}}^2)  \biggr)\biggr],
\end{align}
where $c_\rho=\pi^2g_*(T_{R})/30$, and the equilibrium density $X_{\rm{EQ}}$ is
\begin{equation}
	X_{\rm{EQ}}=\frac{A^3}{T_R^3}\frac{g_{\rm{eff}}Tm_{\rm{DM}}^2}{2\pi^2}K_2\left(\frac{m_{\rm{DM}}}{T}\right),
\end{equation} 
with $K_2$ is the modified Bessel function of second kind.  The dimensionless Hubble parameter is given by 
\begin{equation}
	\tilde{H}=\sqrt{3}  H A^{3/2}T_R^{-2}M_{\rm{Pl}}.
\end{equation}

There are five parameters in our analysis: modulus mass $m_\phi$, the Higgsino DM mass $m_{\rm{DM}}$, the reheating temperature $T_R$, the mass splitting parameters of charginos and neutralinos, i.e., $\Delta m_{\chi^\pm}$ and $\Delta m_{\chi^0}$, respectively. The decay rate of modulus is given by Eq.\ \eqref{Tr} once $T_R$ is given. In addition, the value of $\kappa$ can be inferred from $T_R$ and $m_\phi$ using Eq.\ \eqref{Gammaphi}. We start our numerical calculation from $H=10^{20}\Gamma_\phi$, at which the universe is dominated by the modulus field; the initial density of the modulus is given by 
\begin{align}
	\Phi_I=\frac{3H_I^2M^2_{\rm{Pl}}}{T^4_{\rm{R}}},
\end{align}
where the scale factor is normalized so that the initial value of $A$ is equal to $1$.
The initial densities of Higgsino DM and radiation are neglected. 

We solve the reformulated Boltzmann equations to the epoch well withing the radiation-dominated era; the initial and the final values of $A$ are $A_I=1$ and $A_E$, respectively.  Then, the relic abundance of the Higgsino DM is given by
\begin{align}
	\Omega_{\rm{DM}}h^2=m_{\rm{DM}}\frac{X(T_E)}{R(T_E)}\frac{A_ET_Eg_*(T_E)h_*(T_{\rm{now}} )}{2T_{\rm{now}} T_Rh_*(T_E)}\Omega_\gamma h^2,
\end{align}
where $\Omega_\gamma h^2\simeq 2.473\times 10^{-5}$ is present radiation density and $T_{\rm{now}}\simeq 2.35\times 10^{-13}\ \rm{GeV}$ is today's CMB temperature \cite{ParticleDataGroup:2024cfk}.

\section{Result}
\label{sec:result}

In this section, we present our numerical results of the Higgsino DM relic abundance.  Throughout our numerical study, we take $m_{\phi}=10^5\ \rm{GeV}$.

\begin{figure}
   \centering
   \subfigure[$B_{\rm{DM}}=0.01$]{ 
   \label{b=0.01}
   \includegraphics[width=0.75\textwidth]{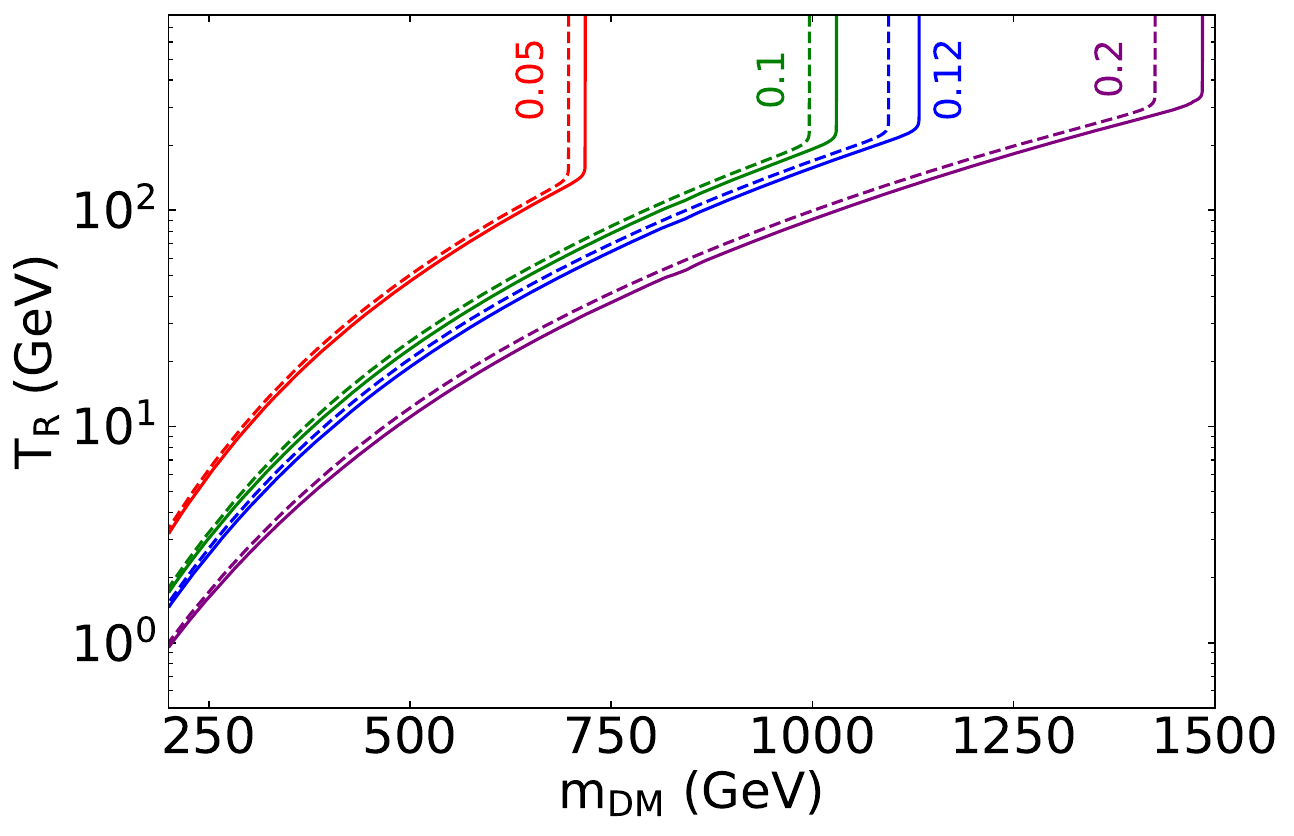}}
    \vspace{3mm}
    \subfigure[$B_{\rm{DM}}=1$]{
    \label{b=1}
     \includegraphics[width=0.75\textwidth]{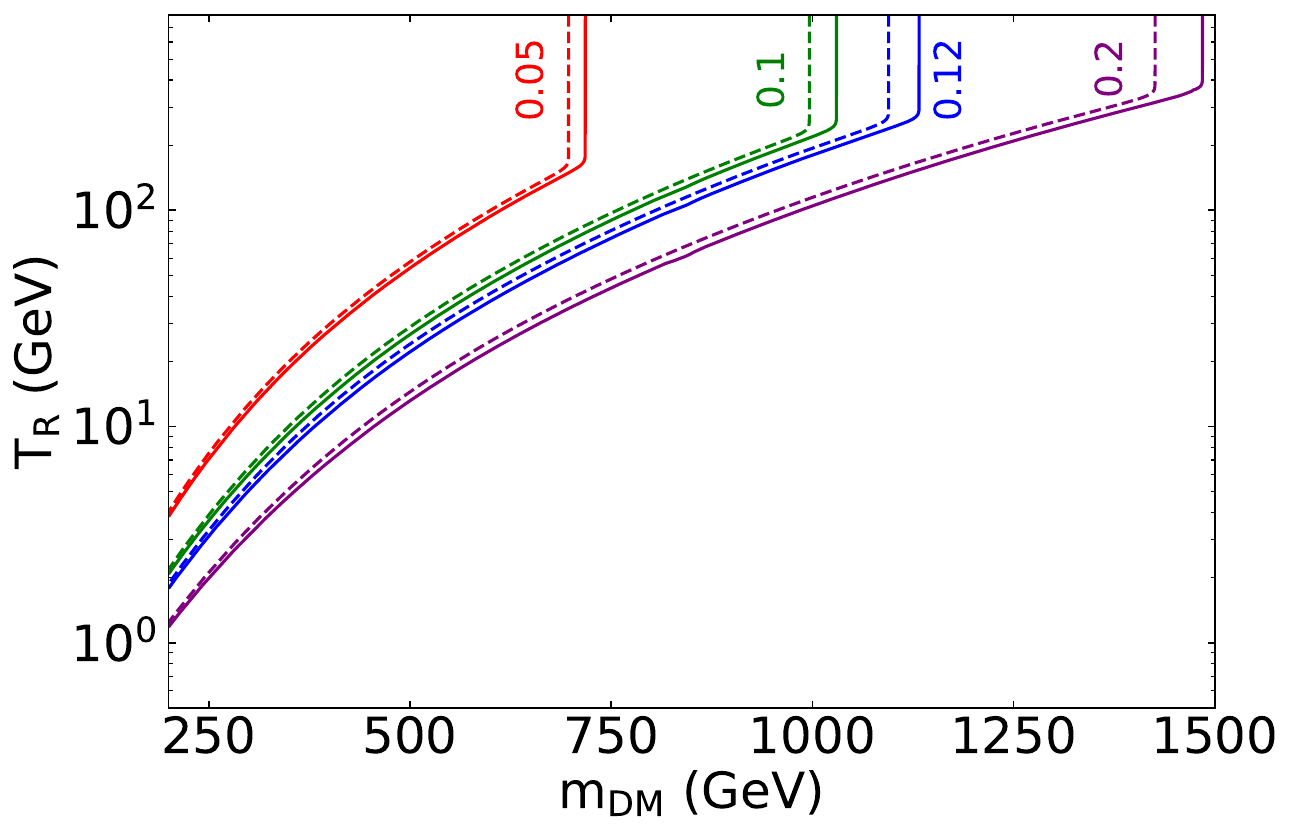}
     }
 \caption{The required reheating temperature vs the mass of Higgsino DM to give relic abundance $\Omega_{\rm{DM}} h^2=0.05$ (red), $0.1$ (green), $0.12$ (blue), and $0.2$ (purple).
The mass splittings are  $(\Delta m_{\chi^\pm}, \Delta m_{\chi^0})=(0.35 \rm{GeV},0.1 \rm{GeV})$, the branching ratio is taken to be $B_{\rm DM}=10^{-2}$ in  (a) and $B_{\rm DM}=1$ in  (b). 
}
   \label{fig:0.35-0.1-relic}
\end{figure}

Figs.\ \ref{fig:0.35-0.1-relic} show the contours of constant $\Omega_{\rm{DM}} h^2$ on $T_R$ vs.\ $m_{\rm{DM}}$ plane, adopting Model A  given in Table \ref{table:dm}.  The dashed lines are the results with perturbative annihilation cross sections, while for the solid lines Sommerfeld enhancements are included.  Due to the Sommerfeld effect, the relic abundance $\Omega_{\rm{DM}} h^2$ is reduced by $5$ to $10\%$ depending on $m_{\rm{DM}}$. In addition, with the values of $B_{\rm DM}$ used in our analysis, the annihilation of Higgsinos produced by the modulus decay is efficient so the final relic abundance $\Omega_{\rm{DM}} h^2$ is insensitive to $B_{\rm{DM}}$. From the contour corresponding to the observed DM density $\Omega_{\rm{DM}} h^2=0.12$, light Higgsino DM in the mass range of $200\rm{GeV} - 1.2 \rm{TeV}$ can be realized with $T_R$ between $1\ \rm{GeV}$ and $200\ \rm{GeV}$. 
As one can see, at low enough reheating temperature, the reheating temperature relevant for the non-thermal Higgsino DM scenario increases as the Higgsino mass becomes larger.  With large enough reheating temperature $T_R$, on the contrary, the relic abundance loses its dependence on the reheating temperature. This is because the standard freeze-out of Higgsinos occurs after the modulus decay, then $\Omega_{\rm{DM}} h^2$ becomes independent of $T_R$.
Note that, for $m_{\rm{DM}}=1\ \rm{TeV}$, for example, the reheating temperature relevant for the Higgsino DM scenario is $\sim 150\ \rm{GeV}$ which is higher than the freeze-out temperature $\sim m_{\rm{DM}}/20$; this is because modulus has not finished decaying at $T\sim T_R$, so Higgsino production from the modulus decay continues until a temperature below $m_{\rm{DM}}/20$. 

We also note here that, with our choices of $B_{\rm DM}$ parameter, the resultant relic density is insensitive to $B_{\rm DM}$ parameter; this is because the annihilation of the Higgsinos is efficient at the time of the reheating. Such a behavior is suggested by Eq.\ \eqref{relic estimate}. As we have fixed $T_R$, changing $m_{\phi}$ only changes the number density of $\phi$ at the reheating. This effectively changes $B_{\rm DM}$ and the result is insensitive either.

\begin{figure}[t]
	\centering    
 \includegraphics[width=0.75\textwidth]{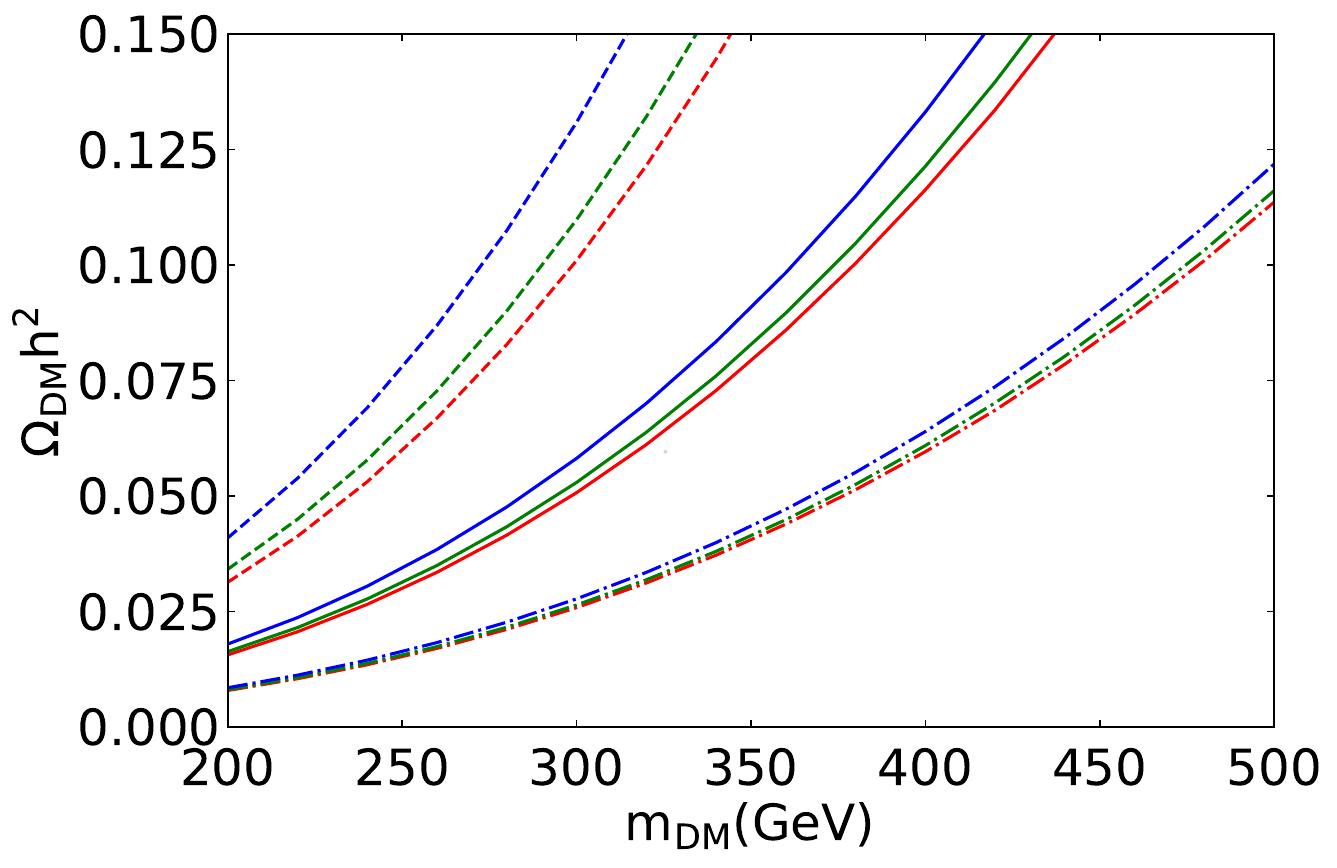}
	\caption{The relic abundance of the Higgsino DM as a function of the DM mass for Model A (red), B (green), and C (blue). The reheating temperature is taken to be 
 $T_{\rm R}=5\ {\rm GeV}$ (dashed), $10\ {\rm GeV}$ (solid), and $20\ {\rm GeV}$ (dash-dotted). Here, we use $B_{\rm DM}=10^{-2}$.}
	\label{fig:relic mass splittings}
\end{figure}
 
In Fig.\ \ref{fig:relic mass splittings}, we show the results of $\Omega_{\rm{DM}} h^2$ as a function of $m_{\rm{DM}}$; the reheating temperature is taken as 
$5\ {\rm GeV}$ (dashed), $T_R=10\ \rm{GeV}$ (solid), and $20\ {\rm GeV}$ (dash-dotted). For such small $T_R$, as one can see from our estimate of DM density Eq.\ \eqref{relic estimate}, $\Omega_{\rm{DM}} h^2$ is proportional to $m_{\rm{DM}}^3/T_R$, 
which can also be inferred from our numerical results shown in Fig.\ \ref{fig:relic mass splittings}. In addition, we can see that the dependence of the relic abundance on the mass splitting parameters is more significant for lower reheating temperature; this can be understood from the fact that, for lower reheating temperature, the Boltzmann suppression for heavier mass eigenstates is more efficient. 

\section{Conclusions and discussion}
\label{sec:conclusion}

In this paper, we have studied the non-thermal production of Higgsino DM via the decay of a late-decaying scalar field. Late-decaying scalar fields show up in a large class of particle-physics models, particularly in supersymmetric extensions of the standard model in which new scalar fields inevitably exist; the examples include moduli fields in string-inspired model, scalar counterpart of the axion in supersymmetric model and so on. Supersymmetric models with $R$-parity conservation provides a natural framework to introduce the DM candidate and the understanding of the production mechanisms of the DM in the early universe are crucial to examine the DM candidates.

We have paid particular attention to the Higgsino LSP scenario, which is one of the popular and well-motivated scenario in the MSSM. We have performed a detailed analysis of the non-thermal production processes of Higgsino DM due to the decay of late-decaying scalar field. We have assumed that superparticles other than Higgsinos are so heavy that they do not affect the relic abundance of the Higgsino DM via coannihilation. We have numerically solved the Boltzmann equation governing the number density of Higgsino LSP, taking into account all the relevant annihilating channels among Higgsinos. Because Higgsinos have gauge quantum numbers and couple to electroweak gauge bosons, a non-perturbative effect, so called the Sommerfeld effect, may be significant; in our calculation, the Sommerfeld effect has been properly taken into account in the calculation of the annihilation cross sections of Higgsinos.

We have shown the relic abundance of Higgsino DM produced by the decay of the late-decaying scalar field.  Contrary to the case of the thermal production, in which the observed DM density is realized with the Higgsino mass of $\sim 1.2\ {\rm TeV}$, the proper DM abundance is realized with lighter Higgsino mass when the reheating temperature is lower than the Higgsino mass. For the case of $T_{\rm R}=5$, $10$, and $20\ {\rm GeV}$, the Higgsino mass realizing the observed DM relic density is about $300$, $400$, and $500\ {\rm GeV}$, respectively. We have also shown that the Higgsino mass relevant for the Higgsino DM scenario depends on the mass splittings among Higgsinos; it changes $10-20\ {\rm GeV}$ as we change the mass splitting parameters.

%%%%%%%%%%%%%%%%%%%%%%%%%%%%%%%%%%%%%%%%%%%%
\acknowledgments
%%%%%%%%%%%%%%%%%%%%%%%%%%%%%%%%%%%%%%%%%%%%

We would like to thank Masato Senami for the useful discussion.
The work of H.F. was supported by JSPS KAKENHI Grant No.\ 24K17042.
The work of Q.L. is supported by the Global Science Graduate Course (GSGC) program of the University of Tokyo.
The work of T.M. is supported by JSPS KAKENHI Grant No.\ 23K22486. 

%%%%%%%%%%%%%%%
\appendix
%%%%%%%%%%%%%%%

\section{Potentials and absorptive parts}

In this appendix, we summarize the potentials and absorptive parts for the Higgsino DM, which are necessary to calculate the annihilation rate with including the Sommerfeld effects.

\subsubsection{$Q=0, S=0$}
The two-body states are 
\begin{align}
    \Phi_0^S=(\chi^+\chi^-,\chi_1^0\chi_1^0,\chi_2^0\chi_2^0).
\end{align}
The potential is 
\begin{align}
\mathcal{H}_{Q=0}^{S=0}=    \Phi_0^{S\dagger} \begin{pmatrix}
        2\Delta m_{\chi^\pm}-\frac{A^2\alpha_2e^{-m_Zr}}{4c_w^2 r} - \frac{\alpha}{r} & -\frac{\sqrt{2}\alpha_2e^{-m_Wr}}{4r} & -\frac{\sqrt{2}\alpha_2e^{-m_Wr}}{4r} \\
        -\frac{\sqrt{2}\alpha_2e^{-m_Wr}}{4r}  & 0 & -\frac{\alpha_2e^{-m_Zr}}{4c_w^2r} \\
        -\frac{\sqrt{2}\alpha_2e^{-m_Wr}}{4r} & -\frac{\alpha_2e^{-m_Zr}}{4c_w^2r} & 2\Delta m_{\chi^0}
    \end{pmatrix}\Phi_0^S.
\end{align}
The absorptive part is 
\begin{align}
    \Gamma_{Q=0}^{S=0} =&\ \Phi_0^{S\dagger}\ \qty(\Gamma_{WW}^{S} + \Gamma_{ZZ}^{S} + \Gamma_{Z\gamma}^{S} + \Gamma_{\gamma \gamma}^{S} )\ \Phi_0^S,
\end{align}
where 
\begin{align}
    \Gamma_{WW}^S =&\ \frac{\pi\alpha_2^2}{64m^2} \mqty(8 & 4\sqrt{2} & 4\sqrt{2} \\ 4\sqrt{2} & 4 & 4 \\ 4\sqrt{2} & 4 & 4), \\
    \Gamma_{ZZ}^S =&\  \frac{1}{c_W^4}\frac{\pi\alpha_2^2}{64m^2} \mqty(4(1-2s_W^2)^4 & 2\sqrt{2}(1-2s_W^2)^2 & 2\sqrt{2}(1-2s_W^2)^2 \\ 2\sqrt{2}(1-2s_W^2)^2 & 2 & 2 \\ 2\sqrt{2}(1-2s_W^2)^2 & 2 & 2), \\
    \Gamma_{Z\gamma}^S =& \frac{s_W^2}{c_W^2}\frac{\pi\alpha_2^2}{2m^2} \mqty((1-2s_W^2)^2 & 0 & 0 \\ 0 & 0 & 0 \\ 0 & 0 & 0), \\
    \Gamma_{\gamma\gamma}^S =& s_W^4\frac{\pi\alpha_2^2}{m^2} \mqty(1 & 0 & 0 \\ 0 & 0 & 0 \\ 0 & 0 & 0).
    \end{align}

\subsubsection{$Q=0, S=1$}
The two-body states are 
\begin{align}
    \Phi_0^{V}=(\chi^+\chi^-, \chi_1^0\chi_2^0).
\end{align}
The potential is 
\begin{align}
\mathcal{H}_{Q=0}^{S=1}=    \Phi_0^{V\dagger} \begin{pmatrix}
        2\Delta m_{\chi^\pm}-\Delta m_{\chi^0}-\frac{A^2\alpha_2e^{-m_Zr}}{4c_w^2 r} - \frac{\alpha}{r} & -i\frac{\alpha_2e^{-m_Wr}}{4r}  \\
        i\frac{\alpha_2e^{-m_Wr}}{4r} & -\frac{\alpha_2e^{-m_Zr}}{4c_w^2 r} 
    \end{pmatrix} \Phi_0^{V}.
\end{align}
The absorptive part is 
\begin{align}
    \Gamma_{Q=0}^{S=1} =&\ \Phi_0^{V\dagger} \qty(\Gamma_{WW}^{V} + \Gamma_{Zh}^{V} + \Gamma_{ff}^{V}) \Phi_0^{V},
\end{align}
where
\begin{align}
    \Gamma^{V}_{WW} =&\ \frac{1}{c_W^4}\frac{\pi\alpha_2^2}{192m^2} \mqty(1 & i[c_{2W}^3+4c_W^2s_W^2c_{2W}] \\ -i[c_{2W}^3+4c_W^2s_W^2c_{2W}] & c_{2W}^2), \\
    \Gamma^{V}_{Zh} =&\ \frac{1}{c_W^4}\frac{\pi\alpha_2^2}{192m^2} \mqty(c_{2W}^2 & ic_{2W} \\ -ic_{2W} & 1),
\end{align}
and
\begin{align}
    \Gamma^{V}_{ff,11} =&\ \frac{\pi\alpha_2^2}{3m^2} \sum_f \frac{N_c}{2}\qty(\frac{c_{2W}}{2c_W^2}(T^3-s_W^2Q)+s_W^2Q)^2, \\
    \Gamma^{V}_{ff,22} =&\ \frac{\pi\alpha_2^2}{3m^2} \sum_f \frac{N_c}{8c_W^4}(T^3-Qs_W^2)^2,\\
    \Gamma^{V}_{ff,12} =&\ i\frac{\pi\alpha_2^2}{3m^2} \sum_f \frac{N_c}{4c_W^2}(T^3-Qs_W^2)\qty(\frac{c_{2W}}{2c_W^2}(T^3-Qs_W^2)+Qs_W^2), \\
    \Gamma^{V}_{ff,21} =&\ \qty(\Gamma^{V}_{ff,12})^*.
\end{align}

\subsubsection{$Q=1, S=0$}
The two-body states are
\begin{align}
    \Phi^{S}_{-}=(\chi^-\chi_1^0, \chi^-\chi_2^0).
\end{align}
The potential is 
\begin{align}
 \mathcal{H}_{Q=1}^{S=0}=     \Phi_-^{S\dagger} \begin{pmatrix}
        \frac{\alpha_2e^{-m_Wr}}{4r} & -i\frac{\alpha_2e^{-m_Wr}}{4r} + i \frac{A\alpha_2e^{-m_Zr}}{8c_w^2 r} \\
        i\frac{\alpha_2e^{-m_Wr}}{4r} - i \frac{A\alpha_2e^{-m_Zr}}{8c_w^2 r} & \Delta m_{\chi^0}+ \frac{\alpha_2e^{-m_Wr}}{4r} 
    \end{pmatrix}\Phi_-^S .
\end{align}
The absorptive part is 
\begin{equation}
    \Gamma_{Q=1}^{S=0} =\ \Phi^{S\dagger}_{-}\Gamma^{S}_{-}\Phi^{S}_{-},
\end{equation}
where
\begin{align}
    \Gamma^{S}_{-} =&\ t_W^2\frac{\pi\alpha_2^2}{8m^2} \mqty(1 & i \\ -i & 1).
\end{align}

\subsubsection{$Q=1, S=1$}
The two-body states are 
\begin{align}
    \Phi^{V}_{-}=(\chi^-\chi_1^0, \chi^-\chi_2^0).
\end{align}
The potential is 
\begin{align}
 \mathcal{H}_{Q=1}^{S=1}=   \Phi_-^{Va\dagger} \begin{pmatrix}
        -\frac{\alpha_2e^{-m_Wr}}{4r} & i\frac{\alpha_2e^{-m_Wr}}{4r} + i \frac{A\alpha_2e^{-m_Zr}}{8c_w^2 r} \\
        -i\frac{\alpha_2e^{-m_Wr}}{4r} - i \frac{A\alpha_2e^{-m_Zr}}{8c_w^2 r} & \Delta m_{\chi^0}-\frac{\alpha_2e^{-m_Wr}}{4r} 
    \end{pmatrix}\Phi_-^{Va}.
\end{align}
The absorptive part is 
\begin{equation}
    \Gamma_{Q=1}^{S=1} =\ \Phi^{V\dagger}_{-}\Gamma^{V}_{-}\Phi^{V}_{-},
\end{equation}
where
\begin{align}
    \Gamma^{V}_{-}=&\ \frac{25\pi\alpha_2^2}{96m^2}\mqty(1 & i \\ -i & 1).
\end{align}

\bibliography{papers}% Produces the bibliography via BibTeX.

\end{document}